\documentclass[useAMS,usenatbib]{mn2e}

\usepackage[dvips]{graphicx}

\def \aap {A\&A }
\def \apj {ApJ }
\def \apjl {ApJL }
\def \nat {Nature }
\def \mnras {MNRAS }
\def \aj {AJ }
\def \planss {P\&SS }

\def \plmult  {$27\ $} 
\def \etal {\textit{et al. }}

\title[Identification Of Circumbinary Planets]{An Algorithm For Photometric Identification Of Transiting Circumbinary Planets}
\author[A. Ofir]{A.Ofir$^{1}$\thanks{E-mail: avivofir@wise.tau.ac.il}\\
$^{1}$School of Physics and Astronomy, Raymond and Beverly Sackler Faculty of Exact Sciences, Tel Aviv University, Tel Aviv, Israel}

\begin{document}

\date{Submitted...}

\pagerange{\pageref{firstpage}--\pageref{lastpage}} \pubyear{2002}

\maketitle

\label{firstpage}

\begin{abstract}
Transiting planets manifest themselves by a periodic dimming of their host star by a fixed amount. On the other hand, light curves of transiting circumbinary (CB) planets are expected to be neither periodic nor to have a single depth while in transit. These propertied make the popular transit finding algorithm BLS almost ineffective so a modified version of BLS for the identification of CB planets was developed - CB-BLS. We show that using this algorithm it is possible to find CB planets in the residuals of light curves of eclipsing binaries that have noise levels of $1\%$ and more - quality that is routinely achieved by current ground-based transit surveys. Previous searches for CB planets using variation of eclipse times minima of CM Dra and elsewhere are more closely related to radial velocity than to transit searches and so are quite distinct from CB-BLS. Detecting CB planets is expected to have significant impact on our understanding of exoplanets in general, and exoplanet formation in particular. Using CB-BLS will allow to easily harness the massive ground- and space- based photometric surveys in operation to look for these hard-to-find objects.
\end{abstract}


\section{Introduction} \label{Intro}
Since a large fraction, and maybe even most, stars form in multiple systems [e.g., Duquennoy \& Mayor 1991], one may wish to investigate to relations between stellar- and sub-stellar multiplicities. Indeed, already \plmult of the known exoplanets are known to revolve around one component of a wide binary\footnote{As of Feb. 2008, from \textit{The Extrasolar Planets Encyclopaedia} at http://exoplanet.eu/} (the so called S-type orbit), and several works [e.g., Mugrauer, Neuh{\"a}user \& Mazeh (2007) and references therein] have already investigated this relation. 

On the other end of the binary separation scale are binaries with periods of a few days, and sometimes shorter than a day. These short-period binaries are compact enough to allow for a theoretical planet to have a stable orbit outside and around both components (P-type orbit) - hereafter circumbinary (CB) planet. These objects are relatively unstudied, and the most serious attempt so far to detect CB planets was the TEP project [Deeg \etal 1998, 2000, 2008, Doyle \etal 2000] which was focused mainly on CM Dra - but none was found.

Typical planetary transit discovery light curves have very low signal/noise ratios so positive detections so far were only obtained by co-adding multiple events by folding the light curve on the correct period. In this paper we will describe a method for the detection transiting CB planets in the light curve residuals of eclipsing binaries (EBs) - in much the same sense that the BLS algorithm [Kov{\'a}cs, Zucker \& Mazeh (2002) - hereafter KZM] is used to find planets in the light curves of single stars.

This paper is organised as follows: in section \ref{Review} we will briefly review the current literature about CB planets. In section \ref{Identifying} we will list the special difficulties one faces when trying to identify transiting CB planets, and give an algorithm that solves most of them. In section \ref{Tests} we will present tests of an initial implementation of the algorithm on simulated data. Finally, in section \ref{Discussion} we will discuss some of the implication the proposed algorithm.

\section{Short Literature Review}
\label{Review}

To our knowledge, so far there have been three announcements of the possible detection of a CB planet - albeit none of which is transiting (in chronological order):

\begin{itemize}
\item Bennett \etal (1999) claimed the detection of a CB planet via microlensing, but Albrow \etal (2000) later found that the light curve can be explained by a binary star where the binary orbital motion had been resolved by the motion of the caustics.

\item Deeg \etal (2000, 2008) claimed that they were able to detected non-linear changes in the observed -- calculated (O-C) eclipse times of M dwarf eclipsing binary CM Dra. They then fitted two models to the data - both including a planetary mass third body orbiting CM Dra. We note that both of these fits critically depend on data from a single epoch and are invalid without it.

\item Correia \etal (2005) had detected radial velocity (RV) variation of HD 202206 consistent with a three-body system. At $m \hspace{3pt}\mathrm{sin}i = 17.4 M_{Jup}$ the inner "planet" is heavy enough to border the planet - brown dwarf (BD) regime. Thus, if $\mathrm{sin}i$ is significantly less than unity, the outer planet may be considered as a CB planet around a stellar-BD binary, but probably not a CB planet orbiting two main sequence stars.
\end{itemize}

From the theoretical side, as early as 1994 Bonnell \& Bate (1994) pointed out that the binary interaction with its (natal) circumbinary disc promotes the disc's fragmentation and the creation of additional companions. They also found that the additional companions will, at least initially, have a nearly coplanar orbit with the original binary. 

Holman \& Wiegert (1999) had found that CB planets can have stable orbits in all binary configurations (i.e., at different mass ratios $q$ and orbital eccentricities $e$) starting at some critical distance $a_{crit}$ and farther out. Their simulations showed that $a_{crit}$ may be as low as 2 (in binary semi-major axis units) for near-circular binaries, and they also found that some configurations may be stable interior to $a_{crit}$ due to resonances. Closeness of CB planets to the host binary is an observationally desired attribute since such planets may complete multiple orbits in the typical survey time span of one observing season.

CB planets are expected to from in CB discs, and indeed several CB discs were already observed [e.g., Duch{\^e}ne \etal (2004), Monin et al.(2007)]. In turn, several theoretical works had investigated the migration and evolution of planets embedded in CB discs (e.g., Pierens \& Nelson 2007, Quintana \& Lissauer 2007 and references therein). These authors too found that CB planets can grow and have stable orbits close to the host binary.

To summarise: simulations show that CB planets can form and survive for long periods even rather close to their host binary, and are more likely to be coplanar with their host binary. This, in turn, gives us some optimism as for the prospects of having transiting planets around EBs.

\section{Identifying transiting CB planets} \label{Identifying}

\subsection{The Problems And The Solutions - General View}

\begin{figure}
\includegraphics[width=0.5\textwidth]{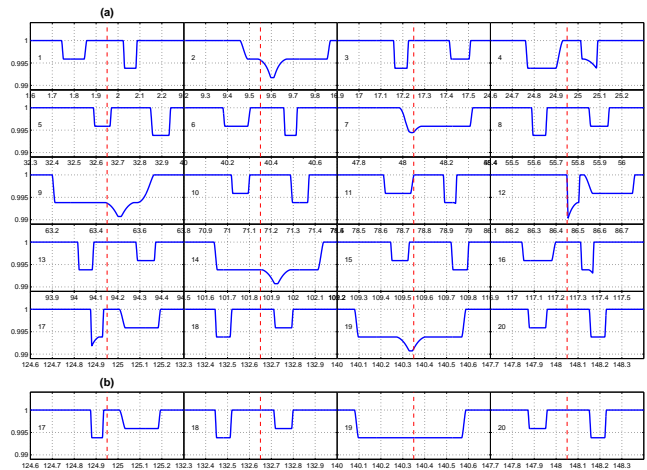}
\caption{panel a: A gallery of 20 consecutive transit events of a system identical to the Default system (see \S\ref{TheTests}) but continuously  sampled. Events are numbered at the left of each sub-panel. Note the highly variable durations and depths: all are purely geometrical effects (no limb darkening). The (red) dashed vertical lines are uniformly spaced between the 1st and 20th transit to show relative shifts of the times of transit. Panel b: The last four transits from panel (a) after regularisation (see \S\ref{CBBLSalgorithm}).}
\label{TransitsGallary}
\end{figure}

Transiting planets manifest themselves by a periodic dimming of their host star. The efficient and popular BLS algorithm [KZM] relies on that periodicity, together with a simple 2-level model of a (low signal/noise) light curve. However, light curves of transiting CB planets are neither periodic nor do they have a single depth while in transit, from the following reasons (see Fig. \ref{TransitsGallary}, panel (a) for illustration):

\begin{enumerate}
\item Photometric characteristics: There is no single amount of dimming of the stellar flux while in transit, since this amount depends on the surface brightness of the hidden part of the binary components relative to the instantaneous total binary flux. Furthermore, the two stars repeatedly eclipse each other (changing the instantaneous total flux) and may be tidally distorted and/or have different surface brightness.

\item Temporal characteristics: The transit signal is not periodic since each time the planet transit one member of the binary - that star is at a different position along the binary orbit, or it may transits the other member altogether. Moreover, transit durations are highly non-uniform: the motion of the binary members - and not of the planet - largely determines the duration of the transits since they can move either in a parallel or anti-parallel to the planet's own motion (enabling very long and very short transits, respectively).
\end{enumerate}

In short, the solutions for the photometric characteristics are: 1) regularising the depths of all transits, and 2) allowing for different effective temperatures of the binary components. The solution for the temporal characteristics is to abandon the view that transits are a function of time: One must recall that transits (and eclipses) are not temporal phenomena, but rather geometrical phenomena - the alignment of celestial bodies.

\begin{figure}
\includegraphics[width=0.5\textwidth]{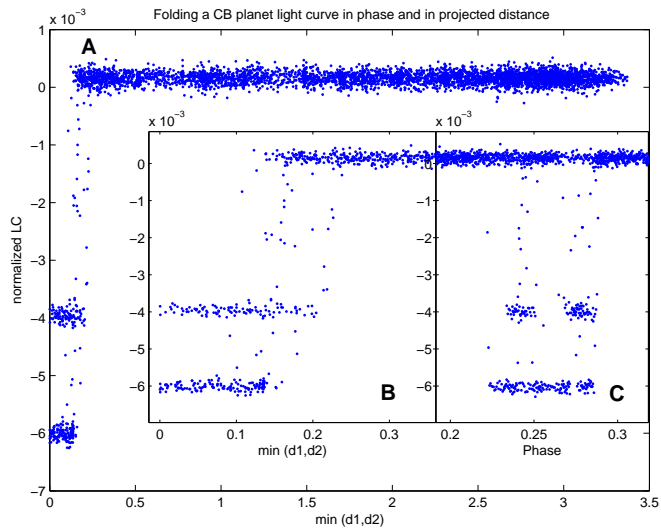}
\caption{Fitting orbital models to the planet allows to "fold" the light curve in projected distance from the members of the binary (labeled as d1, d2). We plot (panel A, also zoomed-in on panel B) such a folding of the Default light curve (see text) with very low white noise of only 0.01\% to aid visibility. The regularised light curve is plotted against min(d1,d2) as derived from one model (only the half of the points where the planet is in front of the stars are shown). Evidently, In-transit points are well-separated from out-of-transit points. The different surface brightness and sizes of the stars mean different depths and distances where transits begin to occurs, respectively. For comparison, a simple phase-folding of the same data is given in panel C showing that in- and out of- transit points are not well separated, significantly reducing the detectability of the signal. Note that for this plot we use only the minimum projected distance - but in CB-BLS more conditions sort out exactly which component is being transited.}
\label{LC-vs-d}
\end{figure}

We remind that in the BLS algorithm, for each test frequency one searches for the phases of the beginning and end of the transit signal in the folded light curve. For CB planets the search is not in time or phase, but rather in orbital parameters space: for a given planetary (and binary) orbit the projected distances between the planet and each of the stars is known, and occurrence of a transit is exactly true or false at each point in time (ignoring planetary ingress/egress). One can then, similarly to BLS, fit a discrete-valued function to the data - where all the in-transit and out-of-transit points are already separated (see Fig. \ref{LC-vs-d}). The output is a multi-dimensional "periodogram" - where each value corresponds to the best fit not only in orbital period, but also all other tested orbital elements.

\subsection{Preparing The Light Curves}  \label{Preparing}

We assume that the all light curves were already searched for periodic variables, but one should take grate care preparing these light curves when trying to identify such a small signal under an already varying background.

\begin{enumerate}
\item All transit searches nowadays use some kind of detrending (such as Syerem [Tamuz, Mazeh, \& Zucker 2005] or TFA [Kov{\'a}cs, Bakos \& Noyes 2005]) in order to reduce the scatter and systematics of the light curves. However, these algorithms expect that all light curves are made of only Gaussian noise + systematics, so the strong EB signal will cause the light curve to be very poorly corrected. Therefore the detrending procedure must be applied on the residuals around some model or smoothed\footnote{We have good experience with the Savitzky-Golay smoothing as implemented in the \texttt{MATLAB} procedure \texttt{smooth}: using it with a large window ($5\%-10\%$ of the phase) and polynomial degree of 3 or 4 usually give excellent results.} light curve iteratively - each time detrending better and generating a better model/smooth. This procedure can have a very significant impact on the quality of the light curve.

\item Once final light curves are obtained, we assume that the binary is accurately solved, probably with the WD code [Wilson \& Devinney 1971, Wilson 1979, 1990] or one of its derivatives, giving: $P_b$, $T_0$, $e$, $\omega$, $i$ and $R_{1,2}$ for binary orbital period, time of priastron passage (or time of primary eclipse for circular orbits), eccentricity, inclination and stellar radii (as a fraction of the binary semi-major axis), respectively.

\item Close binaries are not spherical in shape. CB-BLS will perform better (see \S \ref{CBBLSalgorithm} step (v)) if this is accounted for by pre-calculating the sky-projected shape of the components at each binary phase in the data.
\end{enumerate}

\subsection{CB-BLS} \label{CBBLSalgorithm}

The algorithm we propose, which we dub CB-BLS has the following steps:

\begin{enumerate}
\item The binary model and the data light curve are normalised so that the maximum model flux is exactly 1. The model is then subtracted from the data light curve giving the residual light curve, and for brevity we shall hereafter call that residual light curve just light curve. We will search for a transiting CB planet in this light curve.

\item Regularise the depths: to create a well-defined depth for the transit regardless of the binary eclipses or ellipsoidal variation we multiply the light curve (and the associated errors) with the model at all times, which means that all depths are now well defined as the amount of blocked flux relative to the maximum binary flux. For example, a Jupiter-like CB planet around a binary made of two sun-like stars would create a $0.5\%$ transit at full binary flux, but a $1\%$ transit against full binary eclipse $\sim P_b/4$ later. After regularisation all transits would be $0.5\%$, regardless of binary phase (See Fig. \ref{TransitsGallary} panel (b)).

\item Not having spectroscopic information, we assume a binary mass ratio $q=m_2/(m_1+m_2)$. For each tested $q$ one can now derive the sky-projected relative positions of each component, or specifically: the X and Z position of each component in binary semi-major axis units at each moment of the time series. The coordinates system is set up so that the origin is at the centre of mass, the Y axis is towards the observer, and the YZ plane contains the binary orbital angular momentum vector. A condition for the correct binary mass ratio sampling can be constructed: $\Delta q$ will be such that between two adjacent $q$ values the binary members will change their projected position by no more than one mean stellar radius. Since the length unit is defined as the binary semi major axis then $\Delta q=\textrm{mean}(R)/(1+e)$ where $e$ is the binary eccentricity. We remind that $q$ can be very well constrained from independent sources (spectroscopy) and may need not be searched at all.

\item For each stellar mass ratio we assume a certain planetary orbit. In the simplest case of a circular CB planet with orbit exactly co-planar with the binary orbit, one needs to assume $fp$, $\varphi_0$ and $a$ for planetary orbital frequency, orbital phase at the first data point and planetary semi-major axis (in binary semi-major axis units) respectively. Although it appears that $a$ can be computed from Kepler's laws by $a=(P_p/P_b)^{2/3}$ (where $P_p=1/f_p$), effects of binary-planet interaction cause the effective gravity at the planet, and so the semi-major axis, to be slightly different. One therefore needs to search for a better fitting $a$ in a small range around Kepler's laws value. In addition, $\varphi_0$ also has a natural scale
which can be set similarly to the $q$ condition: $\Delta\varphi_0$ is set so the planet's position will change by less than one mean stellar radius between adjacent $\varphi_0$, or: $\Delta\varphi_0=mean(R)/(2\pi a)$. In practice, one  specifies $a$ as multiples of Kepler's laws value (probably within a small range around unity), and the desired phase resolution as a multiple of above $\Delta\varphi_0$.

While eccentric binary orbits are already accounted for in the current implementation of CB-BLS, Eccentric planetary orbits are not accounted for in the current implementation. Eccentricity is expected to have limited impact since the part of the orbit were transits are possible is fairly small and so the effects of eccentricity will be usually small. The entire search space is then 1+(orbital model) dimensional, or between 4 and 7 dimensional, as long as the low-mass planet approximation in maintained. We note that the original BLS is a 3-dimensional search.

\item At each combination of $q$ and planetary orbit one can compute the projected position of all components, and so determine for each data point whether the planet was transiting either the primary or the secondary binary components. One can then compute the CB-BLS statistic - which is a generalized version of the BLS statistic to a two-box BLS (see below). The best-fitting planetary system will be a peak in that "periodogram" hyperspace, and the exact location of the peak can be found by nonlinear minimisation.
\end{enumerate}

Since the binary components may have different surface brightness the amount of flux blocked by the planet will depend on the binary component being transited. We therefore need to generalize the BLS statistic from a two-level function to a three-level discrete function, namely: $H$ (out of transit) $L_1$ (transit of the primary) and $L_2$ (transit of the secondary). For clarity, we use symbols similar (but not identical) to those in KZM. Let us denote the light curve of $N$ data points by {\{$x_n$\}}, $n=1 \dots  N$, and their respective zero-mean and normally distributed errors {\{$s_n$\}}. The noise is accounted for by assigning weights $w_m=s_m^{-2}[\sum_{n=1}^N s_n^{-2}]^{-1}$. It is further assumed that {\{$w_nx_n$\}} have zero arithmetic average. As explained above, we know exactly which points are in transit in each of the test orbits: {\{$x_i$\}} are all points in transits of the primary star, {\{$x_j$\}} are all points in transits of the secondary star, and {\{$x_k$\}} are all points out of transit. As in KZM, we sum of all weighted squared deviations

\begin{equation}
D=\sum_i w_i(x_i-L_1)^2 + \sum_j w_j(x_j-L_2)^2 + \sum_k w_k(x_k-H)^2 \label{FirstD}
\end{equation}

Minimizing $D$ gives simple arithmetic weights as the best value for each of the levels $L_1$, $L_2$ and \textit{H}: $L_1=s_1/r_1$, \quad $L_2=s_2/r_2$ and $H=\frac{-(s_1+s_2)}{1-(r_1+r_2)}$ Where $s_1=\sum_i x_iw_i$, \quad $s_2=\sum_j x_jw_j$, \quad $r_1=\sum_i w_i$ and $r_2=\sum_j w_j$. Substituting into eq. \ref{FirstD} gives

\begin{equation}
D=\sum_{n=1}^N w_nx_n^2 -\left[ \frac{(s_1+s_2)^2}{1-(r_1+r_2)} + \frac{s_1^2}{r_1}  + \frac{s_2^2}{r_2} \right] \label{SecondD}
\end{equation}

Minimizing D does not depend on the first term of the right hand side of eq. \ref{SecondD}, so the CB-BLS statistics is:

\begin{equation}
CB-BLS=\frac{(s_1+s_2)^2}{1-(r_1+r_2)} + \frac{s_1^2}{r_1}  + \frac{s_2^2}{r_2} \label{CBBLSstst}
\end{equation}

\begin{figure}
\includegraphics[width=0.5\textwidth]{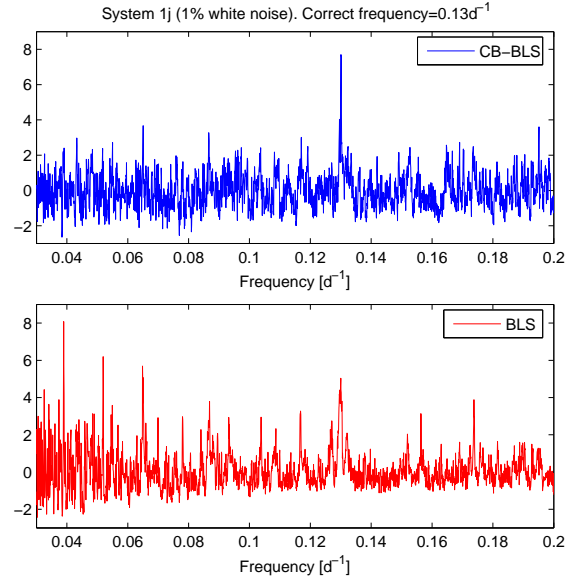}
\caption{An example of the periodograms of CB-BLS (top) and BLS (bottom) for one realisation of System 1j (the Default System with 1\% added white noise). The CB-BLS peridogram was generated as described in \S \ref{TheTests}. The CB-BLS periodogram has no significant aliasing since the model fitted is not periodic. Note that the CB-BLS was created just at the correct sampling limit, meaning that a denser sampling or applying a nonlinear search as described in \S \ref{CBBLSalgorithm} step (vi) will somewhat increase the CB-BLS signal. On the other hand, the BLS signal is already oversampled because of the binning employed.}
\label{OnePeriodogram}
\end{figure}

Its maximum corresponds to the best-fit planetary orbital model. One example of a CB-BLS periodogram compared to a BLS periodogram is shown in Fig. \ref{OnePeriodogram}. It is evident that CB-BLS found the correct frequency with high significance, while BLS did not.

We note that the algorithm will work well even if the third body is a luminous star and not a (dark) planet since this information is already encoded in the binary model as "third-light". We also note that in the current implementation of CB-BLS the CB planet is assumed to be a test mass that does not influence the binary, while exactly this influence is the basis of both eclipse timing and RV. This simplification is a source of noise for CB-BLS.

\section{Tests on simulated data} \label{Tests}

\subsection{Simulated Data Generation}

We integrated the 3D motion of 3 point bodies under mutual gravitational interaction for $\sim$ 150d, and generated simulated light curve for that period (more below). The uniform time steps (50s) were far smaller than the typical exposure time of photometric surveys and continuous so only every eighth simulated point was used to simulate a 400s duty cycle. Next we removed all points meeting modulu(JD) \mbox{$<0.8-sin(\pi \textrm{JD}/150)/8$}, (where JD is the simulated time in days) simulating the lengthening and shortening of nights during the 150d observing season. All the light curves below are therefore almost 9000 data points long. The light curves are generated from the 3D positions using the Mandel and Agol (2002) formalism, so their main limitation is that the stars are assumed spherical. This spherical model will probably be only approximately true for such short-period binaries since they are better described with Roche-lobe geometry. Still, this approximation is well suited for the current implementation of the algorithm since at the analysis stage (when it is determined for each point whether it is in transit or not) it is assumed that the stellar radii are constant. Therefore, the simulated data and the analysis method match in the sense explained in \S \ref{Preparing} step (iii) and \S \ref{CBBLSalgorithm} step (v)). For real data, while preparing the data one can calculate the different projected shapes of the binary components at each binary phase - and thus follow the algorithm (step (v)): at each data point one will still be able to determine whether the planet is transiting one of the binary components or not. In the end, pure Gaussian noise was added to the data.

\begin{figure}
a) \hspace{0.23\textwidth} b)\\
\includegraphics[width=0.23\textwidth]{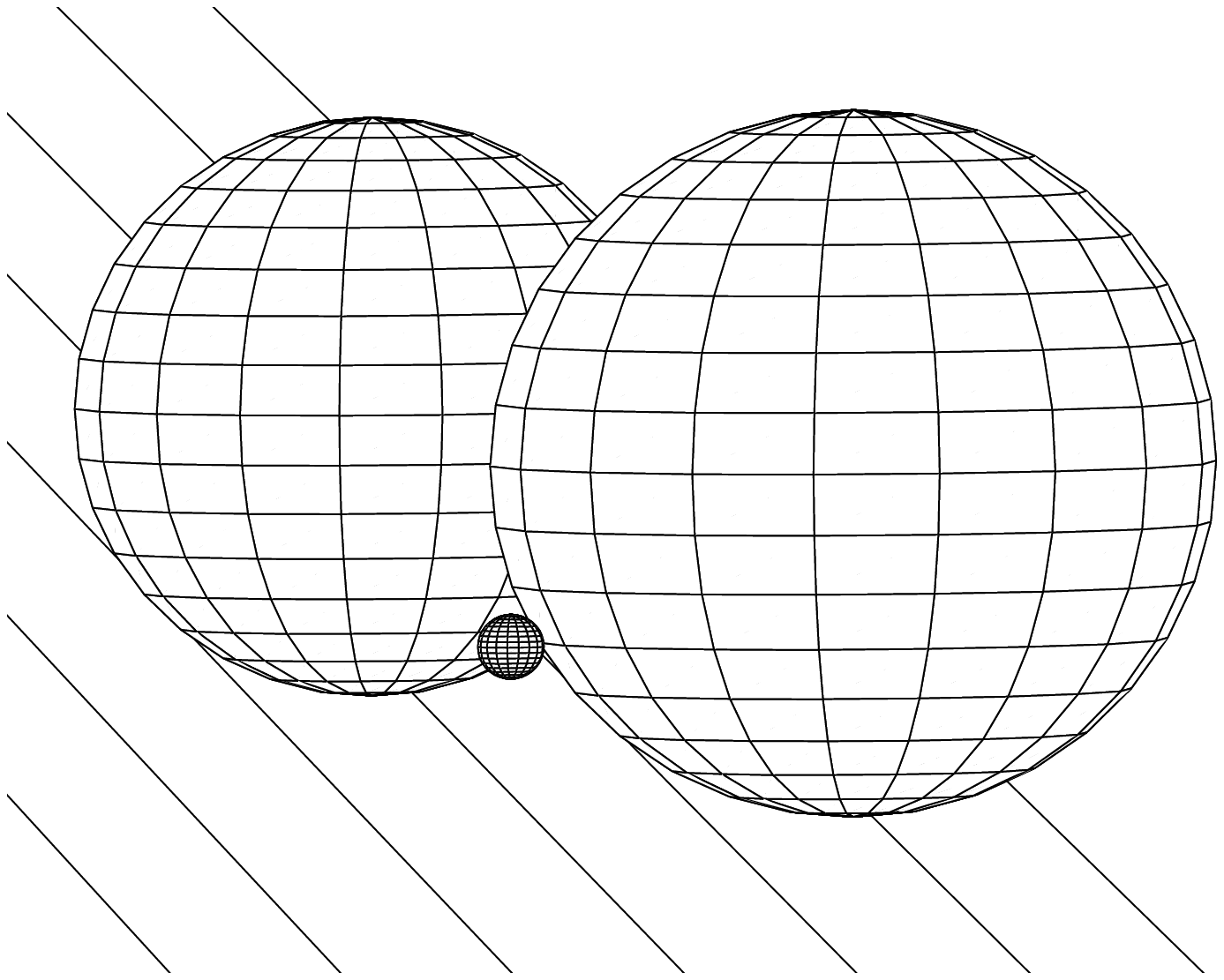}
\hfill
\includegraphics[width=0.23\textwidth]{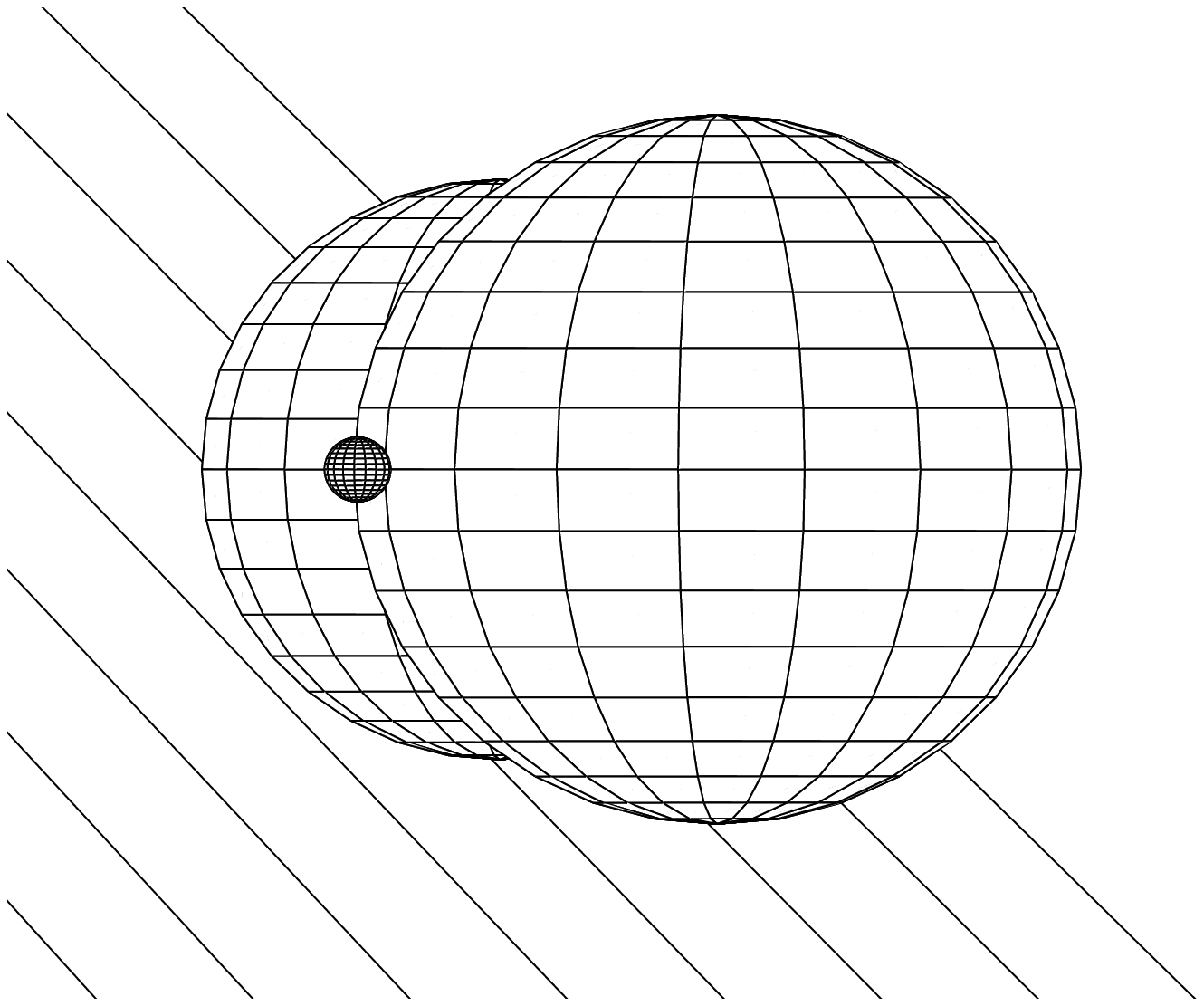}
\caption{Geometrically complex relative positions where it is difficult to account for the contribution of the transit of the \mbox{\textit{occulted}} star. Panel a) Part of the planet's disc transits both stars, and the rest is not transiting either star; we avoided this rare position. Panel b) All the planet's disc transits both stars: in the uniform-disc simulations this case is treated well, but the treatment is only approximated if limb darkening is included}
\label{Positions}
\end{figure}

Since good modeling of EBs is not the topic of this paper, we use the same procedure that generated the data light curve to generate the binary model with exactly the same input values - only with the transit signal excluded from the output. In essence this the perfect model - and modeling errors will indeed be a limitation to CB-BLS (see also discussion on \textsection \ref{Discussion}). We also choose not to simulate limb darkening (although the Mandel and Agol formalism allows for limb darkening) since later, at the analysis stage, these effects should be anyway modelled-out by the binary model and will not add new information on the accuracy of the proposed algorithm. On the other hand, purely geometrical transits allow fewer distracting effects when introducing a new algorithm as in this paper. For a graphical illustration of the effects of limb darkening see Fig. 1 of Deeg \etal (1998). Finally, using the Mandel and Agol formalism prevented us from computing accurately the geometrically complex case of the planet partially transiting both stars simultaneously (see fig. \ref{Positions}, panel (a)). Being quite rare, this configuration was simply avoided in the simulations.

\subsection{Tests}\label{TheTests}

We simulated several systems varying different system parameters in order to explore the properties of CB-BLS. We tabulate the datasets used in this paper in table \ref{SimulatedSystems}, and use the dataset names mentioned there in the following text. Each system was realised 50 times with random white noise, and unless specified otherwise, the results below are the median result of 50 similar realisations. The Default System is the "easiest" since it is very close to the circular, coplaner model this implementation of CB-BLS searches for. The Default System is a binary made from two equal total luminosity stars -- a primary with mass of 1.1M\sun, and radius of 1.1R\sun \space and a secondary with Mass and radius of 0.9M\sun, 0.9R\sun \space (i.e., the latter has higher surface brightness). The system centre of mass is at the origin. The binary orbit is circular with period $P_b=1.23456d$.

The Default CB planet has mass of  0.001M\sun and radius of 0.1R\sun, and a period of $P_p=7.89012d$. The photometric signals are therefore $\sim 0.4\%$ and $\sim 0.6\%$ deep for the transit of the primary and secondary components, respectively. $P_p$ specifies both the distance and velocity of the planet at $T=0$: at that instant a point-mass at the origin with the combined mass of the binary would induce a circular planetary motion with period $P_p$. All three object are on the X axis at initialization, and the planet is on the positive X side and moving towards the observer, so the initial planetary orbital phase $\varphi_0\approx0.75$. Note that the planet does not move in a circular orbit since it is pulled by two moving bodies. 

\begin{table}
\caption{Simulated systems}
\begin{tabular}{l c c c} 
\hline
\label{SimulatedSystems}
System Name 		& Noise [$10^{-3}$] & $P_p$ [d] & $i$ [deg]\\
\hline
Default	&	0&	7.89012 &	90\\
1a, \ldots, 1t		&	1,2,...20 		&	7.89012 &	90\\
2a, \ldots, 2f 		&	10		&	[9 \ldots ,19].89012 		&	90\\
3a, \ldots, 3h 	&	10 		&	7.89012 &	89.75, 89.5, ..., 88\\
\hline
\end{tabular}
\end{table}

Most of the tests below are two dimensional searches on $f_p$ and $\varphi_0$, where the mass ratio is set on it's correct value of $q=0.45$, and the semi major axis factor is at the default value of $a=1$ (for more details on the dependence of CB-BLS on $q$, $a$, and $\varphi_0$ see \S \ref{OtherResults}). We did not try to look for the absolute maximum using non-linear searches.

Search grid density: since in-transit points are only a few percent of the planetary orbit the density of points in the $\Delta f_p$ axis has to be smaller than a few percent of the Nyquist resolution ($\sim 1/ \mathrm{span}$), so we set $\Delta f=1/\mathrm{span}/100$ where span is the time span of the data, $\sim 150$d in the simulations. For the semi-major axis factor $a$ any significant deviation from unity has a physical meaning so a first guess of $a=1$ may not be bad at all. We note that \textit{a} ranges of 0.95 to 1.07 can be derived for several systems simulated by Holman \& Wiegert (1999) (last three columns of their table 4).

The last step is to straighten the periodogram (remove the typical long-periods rise). We iteratively fit a low-order polynomial and clip to $2\sigma$, which finds the "backbone" of the periodogram. The resultant polynomial is subtracted from the periodogram, and the result is divided by the $\sigma$ of the last iteration, effectively making the periodogram values equivalent to the signal detection efficiency metric SDE from KZM. This procedure allows direct comparison between different period-searching algorithms.

The tests begin with measuring the binary period $P_b$ (using AoV [Schwarzenberg-Czerny 1989]) and $T0$ (weighted average of the times during one eclipse weighted by the depth). We then use the model values for the orbital elements and $R_{1,2}$ to complete the CB-BLS inputs about the binary. The frequency range searched for is between $f_p=0.02$ (to allow 3 full periods within the time span) and $f_p=0.2$. The latter is $\sim 4$ times the binary period, which translates to $\sim 2.5$ times the binary semi-major axis -- close to the $a_{crit}$ for a stable orbit for a CB planet (Holman \& Wiegert 1999). We measure the performance of CB-BLS and BLS by a correctness statistic: the fraction of  realisations that had the highest peak of their periodogram within a small region around the correct value.

\subsubsection{Changing S/N}

\begin{figure}
\includegraphics[width=0.5\textwidth]{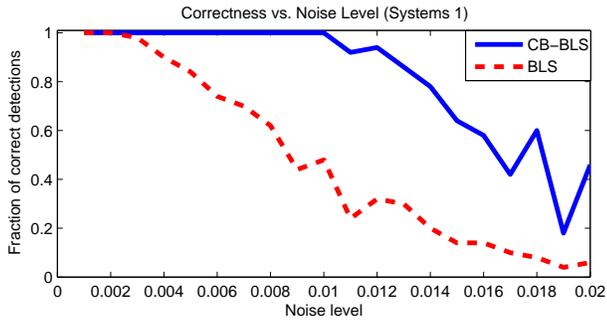}
\caption{For 50 realizations of each of Systems 1a, \ldots, 1t, and for a small region around the correct frequency ($0.129 < f_p < 0.131$), we plot The correctness vs. the amplitude of added white noise for CB-BLS (solid, blue) and BLS (red, dashed).}
\label{SignificanceAndFraction}
\end{figure}

Systems 1a through 1t are the Default System with added white noise with amplitude of $\sigma=0.1\% \ldots 2\%$ respectively. Fig. \ref{SignificanceAndFraction} shows the correctness as a function of the amplitude of the added white noise. Clearly, BLS finds the correct frequency only for the very highest quality light curves - with correct identification falling below $50\%$ already at noise levels of $\sim 0.8\%$, while CB-BLS maintains that performance though noise levels of $\sim 1.6\%$ or more (to be compared with the $\sim0.4\%$ and $\sim0.6\%$ transits). Since in a typical survey there are many more targets with mediocre quality than there are with the highest, millimag-precision targets - the impact on the possible CB planet yield is significant.

\subsubsection{Changing $P_p$}

\begin{figure}
\includegraphics[width=0.5\textwidth]{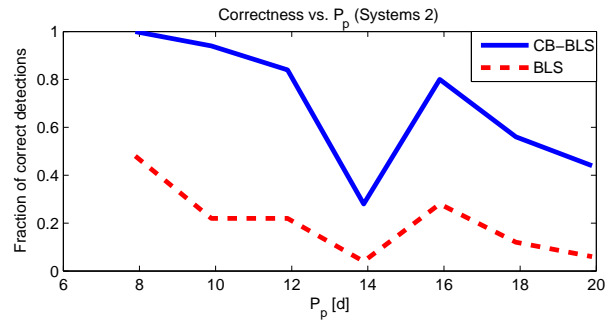}
\caption{For 50 realizations of each of Systems 1j, 2a, \ldots, 2f, we plot the correctness statistic vs. planetary period. The reason for the reduced performance of both BLS and CB-BLS at the $P_p=13.89012d$ is that this period happens to have a particularly bad window function with less than 1/3 of the in-transit points relative to the Default system.}
\label{Systems2}
\end{figure}

Systems 2a through 2f are based on the Default System with $1\%$ added white noise, but with different planetary periods such that for System 2a $P_p=9.89012d$, and each subsequent System $P_p$ is longer by 2 days. These systems, together with System 1j (which has the same noise level and the a 7.89012d period) - are plotted in Fig. \ref{Systems2}. The main effect of the increased period is the reduction of the total number of in-transit points, making the signal detection more difficult. Still, CB-BLS is superior to BLS and maintains a $\sim 50\%$ correct detection ratio all the way to $P_p\approx 20d$.

\subsubsection{Changing $i$}

\begin{figure}
\includegraphics[width=0.5\textwidth]{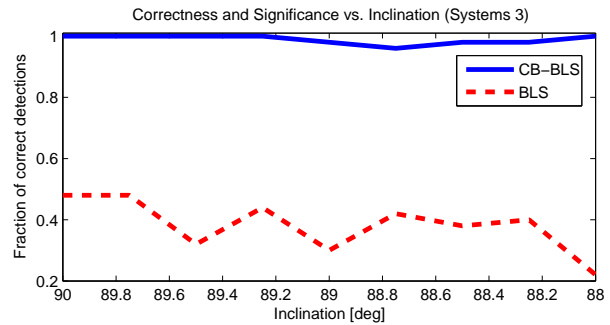}
\caption{For 50 realizations of each of Systems 1j, 3a, \ldots, 3h, we plot the correctness statistic vs. planetary orbital inclination. Note that edge-on is to the left.}
\label{Systems3}
\end{figure}

Systems 3a through 3h are based on the Default Systems with $1\%$ added white noise, only with different inclinations of the planetary orbit between  $90^\circ$ and $88^\circ$ (while transits no longer occur at $\sim 87.5^\circ$ and $\sim 87^\circ$ for the smaller and larger binary components, respectively). The binary star is still exactly edge-on in these systems. Fig. \ref{Systems3} depicts the correctness for these systems and one can see that CB-BLS almost always find the correct period even when the fitted model becomes increasingly inaccurate with decreasing inclination. On the other hand, BLS never finds more than half of the systems.

\subsection{Short Comparison with algorithm by the TEP network}
\label{TDAcomp}

The anonymous referee correctly pointed out that a comparison with the transit detection algorithm used by the TEP project [Doyle \etal (2000) - hereafter the TEP algorithm] is needed. Since a thorough comparison with the TEP algorithm is  beyond the scope of this paper, we only list a few points in lieu:

\begin{itemize}
\item Both algorithms assume for each test orbit a planet in a circular, edge-on and coplanar orbit with the binary.
\item In the TEP algorithm transit events are fitted individually on each night's data, which is reliable only for relatively strong transit signals --- a limitation not present in CB-BLS.
\item The TEP algorithm require multiple fits --- as many fits as there are transit events in each test orbit --- and these will add noise to the detection statistic. CB-BLS requires only a single fit for each test orbit.
\item In the TEP algorithm each test orbit is used to produce a model light curve which is then matched to the data. This requires additional modeling (e.g. planetary radius, limb darkening, etc.) - again adding noise to detection statistic. In CB-BLS no further modeling is required, and the depth(s) are analytically found from the data.
\item In-transit data points carry far more information about the planet than out-of-transit points. However, the TEP algorithm can't handle the longest transits if there are not enough out-of-transit points in a given night (e.g., see events 2,7,9,14 and 19 on fig. \ref{TransitsGallary}). Since these events contain numerous data points and are in part also deeper than other in-transit points, the TEP algorithm therefore gives up on a very valuable portion of the in-transit points. This limitation is not present in CB-BLS. It should be noted, however, that the TEP algorithm was adapted to process a fairly inhomogeneous  data set originating from several different telescopes, and with strong extinction effects due to the red colour of their target star (CM Dra) relative to all available comparison stars.
\end{itemize}

To summarise, the TEP algorithm can be applied to the detection of transiting planets around single stars, and it was already rigorously compared to other such algorithms [Tingley (2003a, and especially 2003b)] and was found to be inferior to BLS. Since CB-BLS and BLS have very similar statistical properties, we believe the TEP algorithm will be inferior to CB-BLS when applied to CB planets.

\subsection{Other Results and Notes}
\label{OtherResults}

We observed the following properties and results of CB-BLS:

\begin{itemize}
\item To show the behaviour of the CB-BLS statistic along the other axes: $q$, $a$ and $\varphi_0$, we used one realization of the Default System with $1\%$ added white noise. CB-BLS produces a very sharp peak on the $\varphi_0$ axis, and the global maximum indeed corresponds to the correct $\varphi_0$, $q$ and $a$ with good precision (see Fig. \ref{QAP-cuts}). Note that no error analysis was carried out and the quoted uncertainties are only a very rough estimate.

\begin{figure}
\includegraphics[width=0.5\textwidth]{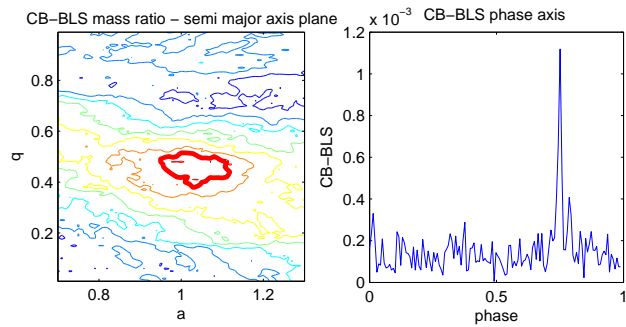}
\caption{Left panel: contours of CB-BLS in the mass ratio -- semi major axis plane. The peak is at $q=0.48^{+1}_{-11}$ and $a=1.07^{+0.04}_{-0.15}$ (where the error ranges are at 90\% the peak level) while the correct values are 0.45 and 0.99, respectively. The nine contours are linearly spaced, so the error ranges are approximately at the boldfaced contour. For illustrative purposes both parameters were sampled at a high resolution of 0.01. Right panel: a cut of CB-BLS in the $\varphi_0$ axis and thought the global maximum. The peak is at the correct value $0.75$.}
\label{QAP-cuts}
\end{figure}

\item CB-BLS takes about 10 times longer to execute (for every value of $q$) than BLS. While BLS can be implemented with a computational shortcut (binning in phase) that shortcut is useless for CB-BLS since the analysis is performed in time, not phase. Therefore, binning the light curve in \textit{time} is indeed possible, but it will cause correlated noise to have increased effect on CB-BLS, and it may cause high dilution of some of the shorter transit events.

\item Since some of the transits of a CB planet are be extremely long we noticed that in the Default System in-transit points make up $3.5\%$ of the data points (Including ingress and egress). Note that this fraction is almost twice as high as the corresponding fraction for a planet around a single star with the same period - alleviating the difficulty of finding planets with longer periods.

\item As illustrated on the left panel of Fig. \ref{QAP-cuts}, the CB-PER statistic is somewhat convoluted on small scales, but the main structures are relatively apparent on large scales. This means that while caution will be needed, sampling CB-BLS on larger scales first (the resolutions recommended and used in this paper) to find the region(s) of promising maxima, and then zooming-in on these regions seems like a good way to perform a more efficient search.

\end{itemize}

We list below a few notes and expectations about CB-BLS:

\begin{itemize}
\item CB planets generate signals in a wide range of durations and depths, possibly reducing effects of red noise [Pont, Zucker \& Queloz  2006] and so false detection rates.

\item By looking at eclipse time variation one may deduce information similar to  RV measurements. In principle, such a signal will allow to obtain all the information available in photometery and RV from the same single light curve. One important consequence of the above is the ability to reduce the false-positives fraction of the final candidates list: 
while CB planets can produce only very small eclipse time variation signal, systems with heavier sub-stellar or low-mass stellar tertiaries may produce much larger signal, and so may be identified as such already from the initial light curve. We comment that such detected systems, with three massive objects in very tight configuration, are interesting systems in their own right. We note that light time effect was not included in the current implementation of CB-BLS and a different analysis tool will be needed for that task.

\item An important source of difficulty for transit surveys is the presence of a number of transit-mimicking phenomena, causing the candidate list to be contaminated by a large fraction of false positives. We believe that CB-BLS will have a lower yield of false positives than searches on single stars from two reasons: firstly, the model we fit is very specific and will probably not fit systems that are not truly three-bodied. Secondly, some of the mimicking systems may be identified as such already from the discovery data thanks to light time effects (see above). However, one should remember the TEP project's warning [Doyle \etal 2000] that when a large number of test orbits are fitted to the data there is a nonvanishing probability that even good candidates may be consequences of random sequences of transit-like noise features.

\item Spectroscopic confirmation of a candidate transiting CB planet is expected to be relatively expensive in telescope time since one should follow the system long enough to allow the three orbits to be disentangled. The lower false positives fraction will allow this high cost to be tolerable.

\item Following up on CB planet may be particularly interesting since these systems have orbital evolution on relatively short time scales (a few 100s of days [e.g., Schneider 1994]), and they produce not one but four distinct Rossiter-McLaughlin effects [e.g., Gaudi \& Winn 2007]: whenever star 1 occult star 2 and vice verse, and whenever the planet transits either one of the stars.

\end{itemize}

\section{Discussion}
\label{Discussion}

We have presented a modified version of the popular BLS algorithm tailored for the detection of circumbinary planets. We have shown that it is superior to BLS for this task, and that using this algorithm it is possible to find CB planets in the residuals of EBs that have noise levels that are routinely achieved by current ground-based transit surveys. 

The algorithm is based on fitting planetary orbits to the data and then applying the CB-BLS statistic. Although more general in principle, we have shown that the simplest edge-on, coplanar and circular model is rather effective as an identification tool (i.e., vs. characterisation). We have shown that CB-BLS maintains high correct identification rate even when the noise level increases, or the planetary period increases, or when the planetary orbital inclination is no longer edge-on.

On the limitations side, the simple fact that the transits need to be discerned against the background of two stars can not be changed. Another possible limitation is EB modeling: since CB planets are found in the residuals of EB light curves, the modeling must be of very high quality to allow the detection of the added weak signal.

Detecting CB planets will have significant impact on the field of extrasolar planets studies as it will expand the possible environments for planet formation significantly. Issues such as migration, stability and planet-disc interaction will have to be further investigated in the context of close binaries. Moreover, since the objects in question are short-period binary stars, detecting such planets may have repercussion on the much more established field of close binary stars (for the closest of which the formation process still not well understood even without the added complexity of planets).

If orbital near-coplanarity is common, then EBs are already pre-selected to be near edge-on. This means that, keeping all things equal, the specific "value" of an EB for the planet hunter is much higher that of a single star. Thus, if one needs to select targets (e.g., the \textit{Kepler} and \textit{CoRoT} space missions) and one aims to find as many planets as possible - one may wish to monitor as many EBs as possible.

So far, no wide-scale searches for transiting CB planets were conducted. We believe the null result of the TEP project is partly due to the fact that it preceded many important advances in the field of transiting exoplanets (the Sysrem, TFA and BLS algorithms to name a few). Later, the same data was used to look for variation of eclipse times minima of CM Dra [Deeg \etal 2000, 2008]. Since in this technique one observes accelerations along the line of sight, it is much more closely related to radial velocity than to transit searches and so is quite distinct from transit searches.

In recent years the rate of transiting planets detection increased dramatically because of the experience gained in doing high relative precision wide-field photometeric surveys. Using CB-BLS will allow to easily harness these massive ground- and space- based surveys to look for transiting circumbinary planets.

\section*{Acknowledgments}

I thank the anonymous referee for the constructive comments that improve the paper. I Acknowledge support from Tsevi Mazeh during earlier stages of work in this field. I Acknowledge using the freely available 
\footnote{http://www.csee.umbc.edu/$\sim$motteler/teaching/courses/parallel\_prog/01a/nbody/nbody.html}
N-body integrator by Howard E. Motteler.

\label{lastpage}

\end{document}